\documentclass{article}
\usepackage{ijcai24}

\usepackage{times}
\usepackage{soul}
\usepackage{url}
\usepackage[hidelinks]{hyperref}
\usepackage[utf8]{inputenc}
\usepackage[small]{caption}
\usepackage{graphicx}
\usepackage{amsmath,amssymb,amsfonts}
\usepackage{booktabs}
\usepackage{algorithm}
\usepackage{algorithmic}
\usepackage{multirow}
\usepackage[switch]{lineno}
\usepackage{natbib}
\usepackage{enumitem}
\usepackage{xcolor}
\usepackage{comment}
\usepackage{subcaption}
\usepackage{pdfpages}

\usepackage{tikz}
\usepackage[edges]{forest}
\definecolor{hiddendraw}{RGB}{205, 44, 36}
\definecolor{hidden-blue}{RGB}{194,232,247}
\definecolor{hidden-orange}{RGB}{243,202,120}
\definecolor{hidden-yellow}{RGB}{242,244,193}

\usepackage{makecell}
\usepackage{colortbl} 
\definecolor{LightGreen}{rgb}{0.8, 1, 0.8} 
\usepackage{listings}

\title{Human Centered AI for Indian Legal Text Analytics}

\author{
Sudipto Ghosh\textsuperscript{1}
\and
Devanshu Verma\textsuperscript{1}
\and
Balaji Ganesan\textsuperscript{2}
\and \\
Purnima Bindal\textsuperscript{1}
\and
Vikas Kumar\textsuperscript{1}
\And
Vasudha Bhatnagar\textsuperscript{1}
\affiliations
\textsuperscript{1}Department of Computer Science, University of Delhi\\
\textsuperscript{2}IBM Research\\
\emails
\{sudipto.mcs22, dverma, pbindal, vikas, vbhatnagar\}@cs.du.ac.in, \\ bganesa1@in.ibm.com
}
\begin{document}

\maketitle

\begin{abstract}
Legal research is a crucial task in the practice of law. It requires intense human effort and intellectual prudence  to research a legal case and prepare arguments. Recent boom in generative AI has not translated to proportionate rise in impactful legal applications, because of low trustworthiness and and the scarcity of specialized datasets for training Large Language Models (LLMs). This position paper explores the potential of LLMs within Legal Text Analytics (LTA), highlighting specific areas where the integration of human expertise can significantly enhance their performance to match that of experts. We introduce a novel dataset and describe a human centered, compound AI system that principally incorporates human inputs for performing LTA tasks with LLMs.
\end{abstract}

\section{Introduction}

\cite{shneiderman2022human} defines Human-Centered AI (HCAI) as a collection of successful technologies that amplify, augment, empower, and enhance human performance.

In the evolving field of human-centered computing, the focus is increasingly on harnessing computing technologies that are not just advanced but also intuitively aligned with human experiences and needs. Within the realm of open-source large language models (LLMs), this perspective becomes crucial as we explore ways to blend the computational might of AI systems with the nuanced understanding and contextual judgment that humans bring to the table.

One particular area that can benefit from HCAI is justice delivery in our court systems. In many countries, the legal system is overwhelmed by a backlog of cases, especially in the lower judiciary. While there are legislations like speedy justice acts, the legal processes are inherently time consuming. AI can help automate legal analytics tasks using Legal Text Analytics (LTA) and speed up justice delivery.

\cite{zaharia2024compound} calls for a shift from models to compound AI systems. We propose creating a human centered, compound AI system comprising of large language models to perform various LTA tasks which deliberately and principally elicit human input.

\begin{figure}
  \includegraphics[width=\columnwidth]{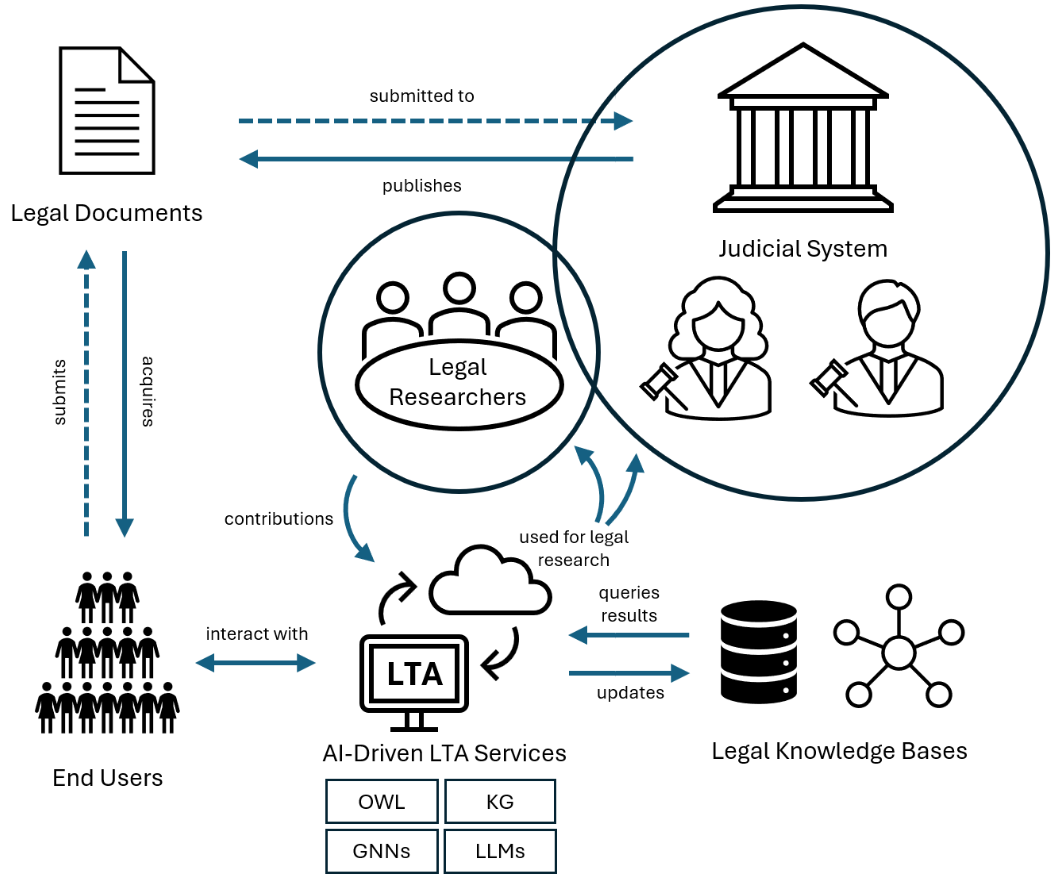}
  \caption{Human computer interaction can bring down information asymmetry in the justice delivery system}
  \label{fig:hcai_lta}
\end{figure}

Figure \ref{fig:hcai_lta} shows a compound AI system, with possible points of interaction between human actors including law professionals and common citizens. Additionally, law students and researchers, petitioners, law activists, etc., also interact with AI-driven LTA services and platforms. Legal experts may use the LTA services for improved efficiency in legal research and help speed up justice delivery. Citizens who are not well versed in legalese can use the services to understand legal documents, do basic research for drafting petitions and be able to submit better responses to the judicial system after taking help from LTA services.

The motivation for a human centered, compound AI system involving models and humans is our observation that the recent boom in Generative AI has not translated to proportionate rise in impactful applications. The reasons for this are low trustworthiness, lack of focus on the common citizens, and general unavailability of resources which deteriorates in domains that directly touch human lives.

In this position paper, we identify few prominent legal text analytics tasks undertaken by legal researchers and self-represented litigants and comment on the performance of existing models on these tasks. We introduce a new dataset and describe how large language models, eliciting human inputs at all levels, can better serve the needs of the people.


Our paper is organised as follows. In Section \ref{sec:related_work}, we survey related work in Legal Text Analytics. In Section \ref{dataset}, we introduce a new dataset for the human centered AI system that we propose. In Section \ref{sec:lta}, we discuss existing problems and how human input can help improve AI systems and in Section \ref{inlegalllama}, we briefly discuss our future work, a large language model to be built incorporating HCAI principles.

\section{Related Work}
\label{sec:related_work}

In this section, we present a non-exhaustive survey of the recent works in tasks relevant for legal text analytics.

\subsection*{Legal Knowledge Graph}
Automatic Knowledge Graph Construction (AKBC) has been popularized since the Knowledge Base Population track \cite{ji2010overview} organized by TAC. Domain specific knowledge graphs (\cite{abu2021domain}) remain an ongoing research area. \cite{dhani2021similar} and \cite{sarika2022constructing} discuss creating legal knowledge graphs using judgements and related documents from Indian courts. The role of human annotations in knowledge graph construction is also a well researched area. \cite{chiticariu2010systemt} proposed a system to extract domain specific entities and relationships from documents. \cite{vannur2021data} discussed fairness in personal knowledge base construction. We can characterize all the methods as based on rule-based or rule-assisted knowledge base construction.

\subsection*{Question Answering on Indian Judgements}
Although question answering is a well-studied problem \cite{pramanik2021uniqorn}, QA systems in the legal domain are not commonly available. Automatic question answering systems not only provide consultancy to litigants who are typically chartering unfamiliar grounds of the legal domain, but are equally beneficial to legal professionals. \cite{ganesan2020anu} presented a question answering system that leverages domain specific knowledge graphs.

\subsection*{Judgment Summarization}
Automatic summarization of judgements, and preparation of \emph{headnotes} (highlighting the point-of-law) help law professionals to locate discussion of a legal issue in lengthy judgements. There have been works in different countries that have addressed this including SALOMON \cite{uyttendaele1998salomon} in Belgium, Letsum \cite{farzindar2004letsum} in Canada, Case summarizer \cite{polsley2016casesummarizer} in Australia. \cite{yizhen2021automatic} summarizes contents of Chinese civil judgments. \cite{kanapala2019text} presented a survey of works in legal text summarization. OpenNyAI \cite{kalamkar-etal-2022-corpus} released annotated Indian court judgments and models for tasks such as automatic structuring of judgments using rhetorical roles and extractive summarization.

\subsection*{Legal Datasets}
\cite{NEURIPS2023_89e44582} introduced LegalBench, a benchmark for measuring legal reasoning in large language models. The Indian Legal Document Corpus published by \cite{malik-etal-2021-ildc} contains 35,000 Indian court judgments and gold standard explanations for the Court Judgment Prediction and Explanation task. We introduce a dataset annotated for tasks like question answering, summarization and petition drafting, which help self-represented litigants to access justice.

\subsection*{Knowledge Infusion}
\cite{chalkidis-etal-2020-legal} introduced the LegalBERT model that continues to be used for tasks on the legal data including our experiments in this work. \cite{paul-2022-pretraining} have introduced InLegalBERT which is trained on Indian legal documents. Infusing knowledge into large language models has been discussed in several works. Two survey papers by \cite{wei2021knowledge} and \cite{yang2021survey} present different methods to infuse knowledge into large language models. \cite{islam2021fair} consumes a knowledge graph for the entity generation task.

\citet{agarwal2020kelm} created a method to translate knowledge graph triples into sentences for enhancing LLM pre-training. \citet{moiseev2022skill} and \citet{agarwal2023there} then directly integrated these triples into LLMs and T5 models, respectively, showing two effective paths for knowledge integration—via natural language or directly from triples. \citet{vasiht2023infusing} took a different approach by using contextual text for embedding knowledge into models.\citet{santos2022knowledge} developed \emph{Knowledge Prompts} for frequent Wikidata entities, refined to aid in triple prediction. \citet{diao2023mixtureofdomainadapters} uses adapters for efficient knowledge infusion into large language models.

\subsection*{Retrieval Augmented Generation}
Text-to-SQL field has seen significant interest with the application of large language models (LLMs) for generating queries. Our focus is on querying case related information for retrieval augmented generation with LLMs. CRUSH4SQL by \cite{kothyari2023crush4sql} employs a retrieval-based method where an LLM generates a simplified schema for query refinement. DIN-SQL by \cite{pourreza2023din} uses a series of prompts to translate natural language into SQL, proving effective on benchmarks like BIRD \cite{li2023can} and Spider \cite{yu2018spider}, surpassing even fine-tuned models.

\subsection*{Resources and Tools}

Services such as \textit{Rocket Lawyer} and \textit{LegalZoom} provide users with the means to create legal documents including pleas, wills, and contracts. \textit{DoNotPay} guides users through various legal processes, offering support from contesting parking tickets to drafting legal documents. Similarly, platforms like \textit{Avvo} and \textit{LawGuru} allow individuals to seek legal advice by posing questions and receiving answers from experienced lawyers, offering invaluable insights for document preparation. Legal aid societies and non-profit organizations like \textit{Pro Bono Net} provide legal assistance support to those with limited resources.

Many court and government websites offer interactive forms and templates to assist self-represented litigants in creating their legal documents with ease. Legal education websites like Nolo expand access to legal information through extensive guides, DIY resources, books, and articles, making legal processes and document preparation more accessible to non-lawyers. Highlighting the intersection of AI and legal aid, \cite{barale2022human} proposes an innovative approach to designing ethical human-AI reasoning support systems for decision-makers in specialized legal areas, such as refugee law, underscoring the potential of AI to augment human capabilities in the legal domain. \cite{joshi2016alda} proposes a method for legal cognitive assistance. 

\section{Dataset}
\label{dataset}
In this section, we describe resources that we have created for enabling legal text analytics. These resources also include human annotations and mechanisms to interact with experts and lawyers. We build on some existing resources from the literature.

\subsubsection*{Legal Knowledge Graph}
\label{legalkg}

A legal knowledge graph can help students familiarize with the legal terms and concepts. Such knowledge graphs can also be used to infuse knowledge into or fine tune large language models (LLMs) to fill gaps in such models where they may not have sufficient domain specific knowledge.

We build on \cite{dhani2021similar} and \cite{sarika2022constructing} to create a legal knowledge graph by scraping the web for court cases, judgements, laws and other cases cited from the judgements etc. In particular, we use court repositories and other public sources in the Indian court system. We further annotate these documents using manually curated dictionaries as described in \cite{vannur2021data}. We process the original documents using Stanza (\cite{stanza}) and extract entities and relations using SystemT (\cite{chiticariu2010systemt}). We also use ground truth labels for citations and similarity from IndianKanoon \cite{ikanoon} and Casemine \cite{casemine}.

We represent the knowledge graph in triples format comprising of subject, object and predicate. Table \ref{tab:akbc} shows the details about the created knowledge graph. This knowledge graph along with other annotations will be made publicly available by us.

\begin{table}[!htb]
\caption{Details of the legal knowledge graph}
\label{tab:akbc}
\centering
\begin{tabular}{lcr}
\hline
\addlinespace
Documents     &{     } & 2,286   \\
Sentences    &{     } & 895,398 \\
Triples &{     } & 801,604 \\
Entities &{    } & 329,179 \\
Relations &{   } & 43 \\
\addlinespace
\hline
\end{tabular}
\end{table}

\subsubsection*{Question Answering}
\label{qa_dataset}

We present a legal question answering dataset \textbf{for law students}, which has been automatically constructed using the \texttt{\small gpt-3.5-turbo} model by OpenAI. We download 45 judgments from the Delhi High Court and extract 1740 paragraphs  containing\textbf{ meaningful} legal text such as paragraphs containing major text as quotation or references from other judgments, names of petitioners, respondents, judges, organizations etc. were discarded. The average paragraph length is 158.91 words. These paragraphs serve as the context for generating the QA dataset.

\begin{figure}[h]
    \centering
    \includegraphics[width=0.8\linewidth]{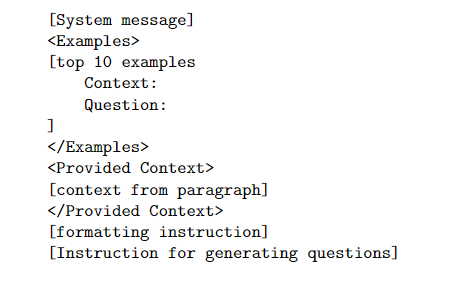}
    \caption{Prompt template for question-answers generation}
    \label{fig:lta-qa-prompt-tem}
\end{figure}

We use the LangChain (\cite{Chase_LangChain_2022}) framework to create a QA generation pipeline. To guide the generation process, we employ few-shot prompting technique, prompt template is shown in figure \ref{fig:lta-qa-prompt-tem}. We use a publicly available short question-answering dataset for the Indian legal system to serve as examples. The dataset comprises of  150 question-answers pairs with questions pertaining to the Indian Constitution, judiciary, legislative, and various socio-political issues in India. For each curated context,  we select 10 question-answers pairs using maximum marginal relevance criterion such that the  questions of the pair are most similar to the  context at hand while maintaining the diversity in these selected questions. The selected QA pairs  and the context are placed into the prompt  (Figure \ref{fig:lta-qa-prompt-tem}) and the model is queried.

\begin{figure}[h]
    \centering
    \includegraphics[width=\columnwidth]{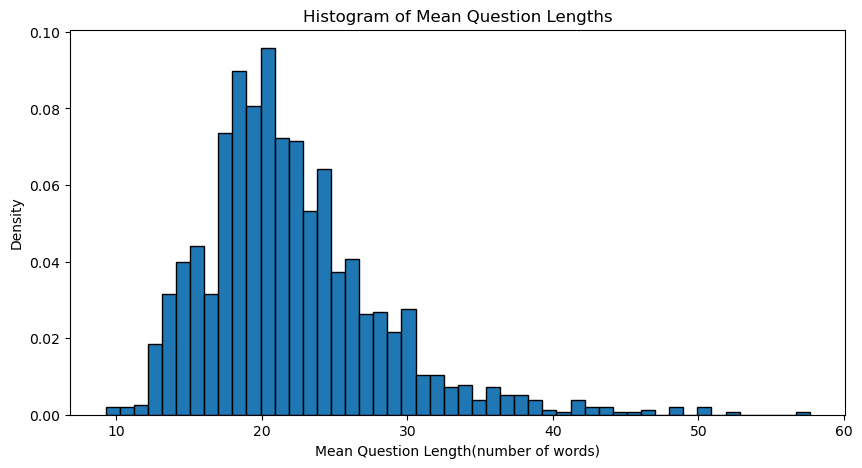}
    \caption{Distibution of question lengths in our Question Answering dataset}
    \label{fig:lta-qa-hist}
\end{figure}

The pipeline  generates the response (questions and corresponding answers) as \textit{JSON} objects. We analyse the length distribution of generated questions and plot it in Figure \ref{fig:lta-qa-hist}. 

\begin{figure*}
    \centering
    \includegraphics[width=.88\textwidth]{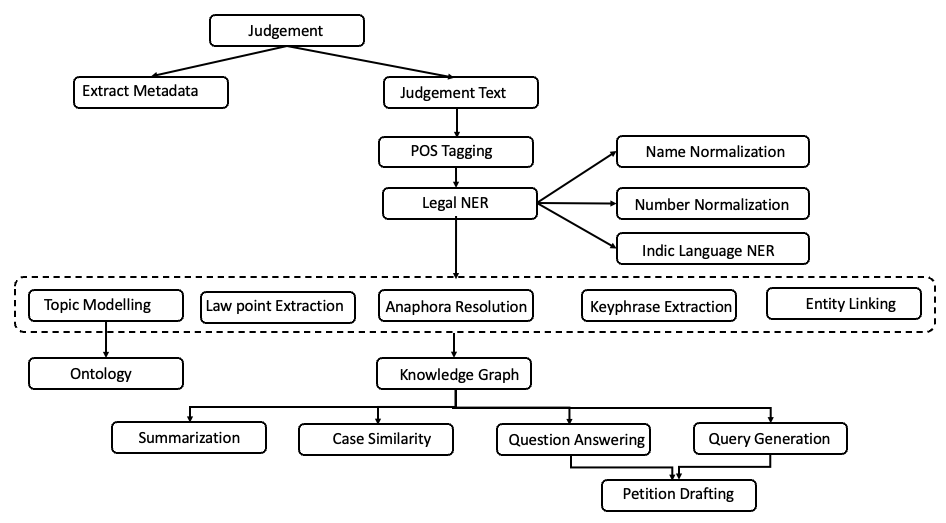}
    \caption{Tasks in Legal Text Analytics}
    \label{fig:lta}
\end{figure*}


\subsubsection*{Text2SQL}
\label{text2sql_dataset}
For the text2sql dataset, that has applications in retrieval augmented generation for petition drafting among other tasks, we extend the question answering dataset described above. We use ideas from \cite{kothyari2023crush4sql} to make the models to hallucinate a database schema required for a query and put them together into a sufficiently large dataset.

\section{Tasks}
\label{sec:lta}

In this section, we present several tasks in Legal Text Analytics that benefit from human interaction either with additional annotations or feedback. We also discuss tasks like petition drafting which is inherently a human activity that draws from peoples' lives. Each of these tasks has been discussed in the literature, where supervised methods have been proposed earlier. We believe annotating samples for training supervised models is not necessarily scalable  for building human-centered AI. Instead, we need to leverage Large Language Models and tailor them to the needs of Legal Text Analytics. With this motivation, we have looked at several tasks, supervised methods currently available for them, and how LLMs can compete and outperform these supervised methods with human intervention. Figure \ref{fig:lta} shows the legal tasks at different levels, from low-level nlp tasks to downstream tasks.

\subsection{Case Similarity}
\label{sec:case_similarity}

\cite{dhani2021similar} [present] a case similarity solution using Graph Neural Networks (GNNs), that can help law practitioners to find similar cases that could lead to early settlements, better case documents, and faster judgements. Following \cite{vannur2021data}, the authors construct a legal knowledge graph using human curated dictionaries to extract entities and relations from documents. The judgements and acts used in that work are from IndianKanoon \cite{ikanoon}, a search engine API and Casemine \cite{casemine}, a resource that provides similar judgements among other features. The authors report manually annotating each judgement as containing a law point or otherwise. They use these annotations as 27 features with one-hot encoding. 

Like in \cite{dhani2021similar}, we extract entities and relations from legal documents, using methods described in \cite{chiticariu2010systemt}, \cite{ganesan2020anu} and \cite{stanza}. This case similarity dataset has 2286 legal documents  with citations from IndianKanoon \cite{ikanoon} and similar cases as recommended by Casemine (\cite{casemine}).

Compared to the vanilla RGCN \cite{schlichtkrull2018rgcn} baseline, they report that the law points identified by legal experts, when used as handcrafted features in addition to the features provide better results. They also compare their method with an RGCN version where the node features are encoded with LegalBERT \cite{chalkidis-etal-2020-legal}.

We conjecture that this task can be performed with Large Language Models, by presenting similar and dissimilar cases, and asking it to predict if two judgements are similar or otherwise. We use the LLaMA-2 70B Chat model and do few-shot prompting, taking a few examples comprising of pairs of document excerpts. We take 1,313 similar pairs from the dataset used by the authors in \cite{dhani2021similar} and an equal number of random pairs. We then ask the LLM if a pair is similar by giving a one word response. We prompt the language model with these 2,626 target pairs of document excerpts taken from the 958 unique judgments in the dataset.

\begin{table}[htb]
    \centering
    \begin{tabular}{lcc}
        \toprule
        \textbf{Model}  & \textbf{Case Similarity} \\
        \midrule
        RGCN baseline & 0.513 \\
        \addlinespace
        RGCN + handcrafted features & 0.556   \\
        \addlinespace
        RGCN + LegalBERT & 0.550 \\
        \addlinespace
        LLaMA-2-70B-Chat & \textbf{0.566} \\
        \bottomrule
    \end{tabular}
    \caption{Performance of different models using ROC-AUC scores}
    \label{tab:rgcn_results}
\end{table}

The case similarity output from the model also has the reasoning behind the Yes-No response and contained 37.62 tokens on average. The performance of the LLaMA-2 chat model (\cite{Touvron2023Llama2O}) is compared with the supervised RGCN model on the Case Similarity task as shown in Table \ref{tab:rgcn_results}. We report the ROC-AUC scores as in \cite{dhani2021similar}. Their model with handcrafted features performed better than the vanilla version with 0.556 ROC-AUC for case similarity. Encoding the features with LegalBERT did not seem to improve performance. LLaMA-2-70B-Chat yields a ROC-AUC score of 0.56626 on the same task.

\subsection{Judgment Summarization}
\label{sec:j-summarization}
Judgement summarization is an important task in legal  research. Since judgments  typically run into tens of pages, stakeholders including law professionals,  activists and common public, require the gist of the document to make sense out of it. Generic and abstractive summarization of judgments is typically useful for common citizens and law students, who are not familiar with legal jargon and prefer summaries in a language that is understandable. Contrastingly, law professionals and legal researchers require aspect-based summaries that meet their personal information needs. Law professionals usually prefer inclusion of legal terms in the summary, thereby necessitating extractive summaries  covering the research  dimension that is important for the law professionals at that point of time.  Thus both approaches, \textit{abstractive} and \textit{extractive}, for judgement summarization are essential and require considerable attention for adoption of AI-driven LTA services. 

Recent advancements in LTA have led to introduction of several state-of-the-art extractive judgment summarizers. \cite{bhattacharya2021incorporating}  propose DELSumm, an extractive summarizer that systematically infuses domain expertise, elucidating the essential information that should ideally be present in the summary of a judgment. \cite{bindal2023citation} summarize landmark judgments based on the references in  citing judgments. Recently,  \cite{dal2023legal} used state-of-the-art LLM GPT4 with prompt engineering to generate extractive summary and report human evaluation results. Extractive summarizers are comprehensible to seasoned professionals as legal judgments feature long sentences and intricate legal terminology.

Abstractive summarization for legal judgements has been practiced using several approaches. Legal-LED and Legal-Pegasus are fine-tuned versions of pre-trained language models LED and Pegasus respectively. Both models are fine-tuned on publicly available legal data from the American judicial system.  \citet{Shukla2022, bindal2023citation} report evaluation results for  summarization of same set of Indian legal judgements using semantics-based metrics and ROUGE scores. \cite{dal2023legal} report human evaluation of abstractive summaries for Italian legal documents using GPT3 and GPT4. \cite{feijo2023improving} generate abstractive summaries by chunking the source text, training summarization models to generate independent versions of summaries, and applying entailment module to mitigate hallucination. The method is evaluated for Brazilian Supreme Court Rulings using ROUGE metric. 

We delve deep into the results reported in \cite{Shukla2022} and \cite{bindal2023citation} for Legal-LED and Legal-Pegasus on the IN-Ext dataset published by  the former. The ROUGE scores for the two semantics-based metrics are reasonably good and a end-user may not suspect any risk in believing the summaries. However an expert analysis of a summary with highest semantic similarity shows a number of problems in the summary.

The cutting-edge models for abstractive judgment summarization, which might have seen some text as part of training data, tend to generate it verbatim in the summary overlooking the context. Resulting inaccuracies, alterations in  proper nouns, locations and numbers lead to significant deviations  from the original content, which is unacceptable in judgment summaries. \citet{dal2023legal} explicitly caution  that abstractive summarization may pose the risk of misleading the readers by generating content absent in the original document. Even a subtle shift in the order of a single word can alter the meaning, as exemplified by the stark difference between "\textbf{accepted} an appeal that had been \textbf{denied}" and  "\textbf{denied} an appeal that had been \textbf{accepted}" \citep{feijo2023improving}. Superiority of \textit{extractive} legal summaries over \textit{abstractive} ones  is well established in some recent studies\citep{Shukla2022, feijo2023improving, dal2023legal, bindal2023citation}.  

Despite its current caveats, moderating research and development of abstractive techniques for judgment summarization is myopic.  We argue that \textit{abstractive} approach  has strong potential for making the long judgment intelligible to a lay person. Ergo, development of reliable and trustworthy summarization methods that unscramble the complex legal language is fundamental for democratizing legal knowledge and ease of access to justice.

Judiciously integrating  knowledge graph, legal dictionary, ontology and other external knowledge sources  with LLMs can not only alleviate introduction of foreign entities and facts in the summary, but also unravel long and complex legal concepts in judgments. Recognizing the  prowess of SOTA LLMs for generating confident, yet simple language, we press for blending sanitation  strategies for  fact-hygiene  and paraphrasing the legal concepts to generate \textit{lay-summaries} of the judgments. Human-centred abstractive summarization of legal judgments is vital for societal good and improved legal-awareness in public. 


\subsection{Petition Drafting}
Petition drafting - the LLM should ask questions that the petitioner can answer. These additional information should strengthen the petition.

\begin{figure*}
    \centering
    \includegraphics[width=0.9\textwidth]{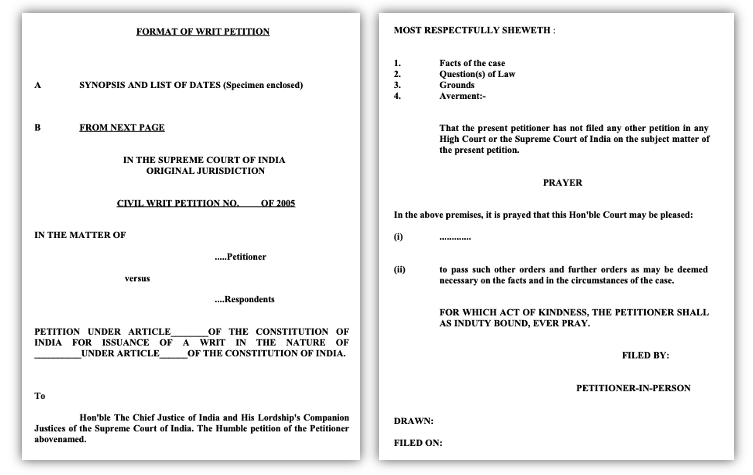}
    \caption{Format of a Writ petition to be filed in the Supreme Court of India. An LLM based solution for assisting self-representing litigants should elicit information to draft such a petition.}
    \label{fig:wp}
\end{figure*}

Petition drafting is a task, that is inherently human centered, especially in the context of Indian court system. Indian courts have the concept of Public Interest Litigations (PILs), using which, any citizen can approach a court of law to seek relief on issues concerning the people. There are, of course, much larger number of people approaching the courts seeking redressal of their grievances.

Enabling people or their lawyers to write well-written petitions can go a long way in getting them access to justice. Given the backlog and the volume of petitions disposed by courts in India and in many other countries, poorly written petitions can add significant cost to both individuals and the society as a whole. Among other things, poorly written petitions could be those that leave out important pieces of information, addressed to the wrong courts or authorities, risk being dismissed as frivolous when infact they are not. Figure \ref{fig:wp} illustrates the format of a writ petition to be filed in SCI.

We propose using LLMs to identify missing information in a petition. This is a qualitatively a much harder task than writing petition which focus on the writing style and presentation. Our task involves making LLMs to identify missing information that should typically be present in the petition. This can be designed as a conversational question answering task. This is closely related to the factuality related work in LLMs, since we do not want the model to ask trivial questions. The model needs to be able to identify salient information in a petition and prompt the user to furnish any missing information.

For example, in a petition about a missing person which is a very sensitive but important judicial function, the petition is expected to provide the time when the person was last seen by a member of the public or a CCTV camera. While we expect this to be a multi-turn conversation similar to \cite{trivedi2023interleaving}, we currently focus on putting together a question answering dataset and evaluating our LLaMA-2 model on the question answering task.

\subsection{Question Answering}
\label{sec:qa}

\begin{table}[!ht]
    \centering
    \resizebox{\columnwidth}{!}{
        \begin{tabular}{lrrr}
            \hline
            \addlinespace
            \textbf{Model} & \textbf{Hits@1 $\uparrow$} & \textbf{Hits@5 $\uparrow$} & \textbf{Hits@10 $\uparrow$} \\
            \addlinespace
            \hline
            \addlinespace
            Legal-BERT-base-uncased & 0.000 & 0.000 & 0.005 \\
            \\[0.05cm]
            InLegalBERT-ft-corpus & 0.005 & 0.005 & 0.015 \\
            \\[0.05cm]
            InLegalBERT-ft-triples & 0.225 & 0.350 & 0.395 \\
            \\[0.05cm]
            LLaMA-2-34B-Instruct & 0.520 & 0.556 & 0.617\\
            \addlinespace
            \hline
        \end{tabular}
    }
    \caption{Comparison of model performance on relation/tail prediction, on a subset of the Legal KG triples}
    \label{tab:expr_cbp_lkg}
\end{table}

Question Answering is an important application to impart legal knowledge to students and answer questions of self-represented litigants in the court system. Considering large language models are often evaluated on the question answering task, they can be used in this legal context too.

\cite{vasiht2023infusing} compare the performance of an InLegalBERT model trained on Indian judgments and additionally fine-tuned on the corpus, with LegalBERT which hasn't seen the corpus. We report their numbers for reference and compare the performance of a LLaMA-2 model on the task, with and without human inputs.

The dataset consists of 4129 question answer pairs from our Indian court judgements dataset. This dataset can be used for different question answering tasks including closed-book QA tasks where a model is expected to answer the question without any further context or external knowledge. We believe models created to assist in petition drafting should be able to do well in this task. We expect such models to be pre-trained with legal documents or triples or fine-tuned as appropriate. In-context learning where the external knowledge is provided as triples or as text \cite{vasiht2023infusing} can also be considered.

However, currently we do not have models infused with legal judgements that can perform this closed-book question answering. We expect most LLMs to perform on the reading compression task but we do not believe that capability is particularly useful for the tasks in Legal Text Analytics. In petition drafting, we need the model to analyze the petition and ask question to fill missing information, like the date on which a particular judgement was delivered.

We believe providing knowledge graph triples or results from a SQL query to be more promising approaches. Our question answering dataset described in Section B of the Supplementary Material gives examples of question answer pairs along with related knowledge graph triples that have been extracted from a judgement. We expect the user to be able to upload a judgement or point to a URL, after which we can generate these triples to be provided as context.

\begin{figure}
    \centering
    \includegraphics[width=\columnwidth]{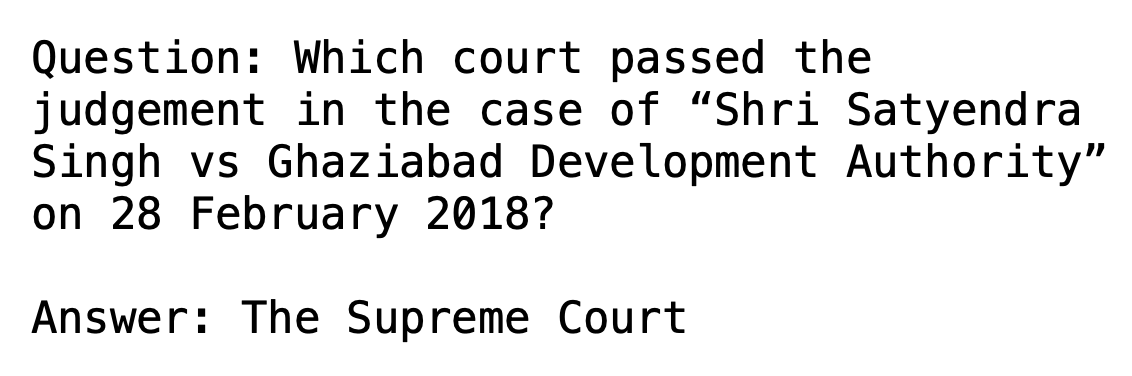}
    \caption{Example Question and Answer in our Indian Court Judgements dataset.}
    \label{fig:qa_example}
\end{figure}

\subsection{Text2SQL}
\label{sec:text2sql}

In the context of legal text analytics, we consider Text2SQL to be predominantly useful for petition drafting and legal research, though other analytics use-cases exist. Generating relevant information about past cases can be helpful to cite judgements, as well as establish facts in a written faction. For example, the date on which a particular judgement was delivered is not part of citation prediction datasets, where as it is required while writing a petition. We propose a retrieval augmented generation (RAG) solution to this problem, where relevant facts of a case can be queried using a Text2SQL model.

\begin{figure}
    \centering
    \includegraphics[width=\columnwidth]{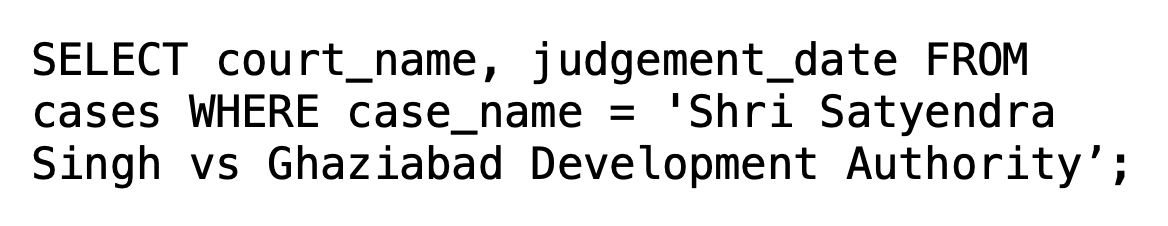}
    \caption{Example SQL query to fill details of a citation in a Writ Petition}
    \label{fig:sql_example}
\end{figure}

However, the resources for enabling such a feature currently do not exist in the literature. We've created a relational database schema consisting of tables and relationships between them as provided in Section C of the supplementary material. Given the question answering dataset described in Section \ref{sec:qa}, and the database schema we have created, we generate text to SQL examples as shown in Figure \ref{fig:sql_example}.

\section{InLegalLLaMA}
\label{inlegalllama}

In Section \ref{sec:lta}, we described the tasks in Legal Text Analytics, that are unique to the legal domain in countries like India. Based on a survey of the literature, and our observations on the performance of LLMs, we propose a large language model infused with legal knowledge and a composite system that elicits knowledge from human users as a better solution for the problems in this space.

Recent work like LegalBERT (\cite{chalkidis-etal-2020-legal}), CaseLawBERT (\cite{zhengguha2021}) and JuriBERT (\cite{douka-etal-2021-juribert}) show the sustained interest of researchers in using language models for downstream legal tasks. However, these models are typically trained on European legal documents, which are structured by nature, and do not perform well in the Indian context directly where courts do not follow standardized structures when publishing legal documents. Under these circumstances, existing PLMs do not work well out of the box and need additional training on local corpora. Works such as InLegalBERT and InCaseLawBERT (\cite{paul-2022-pretraining}) involve training the base models on Indian legal documents and achieve reasonable performance on certain tasks like legal statute identification, semantic segmentation and court judgment prediction.  In Table \ref{tab:rgcn_results}, we observe that open-domain LLMs have comparable performance to RGCN models. Much work still needs to be done to make LLMs useful in LTA tasks that need human expertise.

We conjecture to pre-train a LLaMA-2 foundation model (\cite{Touvron2023Llama2O}) on Indian legal domain corpora and instruction-tune it for a selected set of tasks in the legal domain using a concept-enhanced pre-training objective of entity-concept prediction as proposed in \cite{wang2024concept}, which might help to mitigate hallucination in domain tasks, which could have societal consequences. We aim to fine-tune the model using parameter-efficient fine-tuning methods on feasible domain tasks relevant in the Indian context, and compare its performance on multiple datasets in tandem with other state-of-the-art models, with and without fine-tuning on domain corpora.

\cite{vasiht2023infusing} propose infusing knowledge into LLMs as a general purpose way to improve model performance on documents. They use contextual text in lieu of knowledge graph triples. We propose to use a similar approach to add context to prompts when training the model on Indian legal documents. Using soft prompts and domain concepts in the training process, we will come up with a family of LLMs infused with knowledge about the Indian legal system.

\section{Conclusion}
Our observations of the existing works and applications in the domain of legal text analytics makes it amply clear that we need to develop and deploy human-centered compound Legal AI systems like the one we have described in this position paper. By principally eliciting human input, AI systems can improve the performance of large language models (LLMs) and help our communities build impactful applications, that enable common people to access justice in our court systems.

\clearpage
\bibliographystyle{named}
\bibliography{ijcai24}

\end{document}